\pdfoutput=1
\documentclass[a4paper]{spie} 
\usepackage{amsmath,amsfonts,amssymb}
\usepackage{graphicx}
\usepackage[colorlinks=true, allcolors=blue]{hyperref}
\usepackage{textcomp}

\usepackage{fontawesome}

\usepackage{subfig} % for the subfloat figures.

\usepackage[dvipsnames]{xcolor}
\title{TiEMPO: \\Open-source time-dependent end-to-end model for simulating ground-based submillimeter astronomical observations}

\author[a,b]{Esmee~Huijten}
\author[a,b]{Yannick~Roelvink}
\author[a]{Stefanie~A.~Brackenhoff}
\author[c]{Akio~Taniguchi}
\author[c]{Tom~J.L.C.~Bakx}
\author[d]{Kaushal~B.~Marthi}
\author[a,b]{Stan~Zaalberg}
\author[a,e]{Jochem~J.A.~Baselmans}
\author[f,g]{Kah~Wuy~Chin}
\author[e]{Robert~Huiting}
\author[a,e]{Kenichi~Karatsu}
\author[a,e]{Alejandro~Pascual~Laguna}
\author[c]{Yoichi~Tamura}
\author[h]{Tatsuya~Takekoshi}
\author[i]{Stephen~J.C.~Yates}
\author[a]{Maarten~van~Hoven}
\author[a,b,*]{Akira~Endo}
\affil[a]{Faculty of Electrical Engineering, Mathematics and Computer Science, Delft University of Technology, Mekelweg 4, 2628 CD Delft, the Netherlands.}
\affil[b]{Kavli Institute of NanoScience, Faculty of Applied Sciences, Delft University of Technology, Lorentzweg 1, 2628 CJ Delft, The Netherlands.}
\affil[c]{Division of Particle and Astrophysical Science, Graduate School of Science, Nagoya University, Aichi 464-8602, Japan.}
\affil[d]{Kapteyn Astronomical Institute, University of Groningen, P.O. Box 800, 9700 AV Groningen, The Netherlands}
\affil[e]{SRON---Netherlands Institute for Space Research, Sorbonnelaan 2, 3584 CA Utrecht, The Netherlands.}
\affil[f]{National Astronomical Observatory of Japan, Mitaka, Tokyo 181-8588, Japan.}
\affil[g]{Department of Astronomy, School of Science, University of Tokyo, Bunkyo, Tokyo, 113-0033, Japan}
\affil[h]{Institute of Astronomy, Graduate School of Science, The University of Tokyo, 2-21-1 Osawa, Mitaka, Tokyo 181-0015, Japan.}
\affil[i]{SRON---Netherlands Institute for Space Research,  Landleven 12, 9747 AD Groningen, The Netherlands.}

\authorinfo{Further author information: A.E.: E-mail: a.endo@tudelft.nl}

\begin{document} 
\maketitle

% Abstract <200 words 
% https://www-spiedigitallibrary-org.tudelft.idm.oclc.org/#divPartsofaManuscript
\begin{abstract} 
The next technological breakthrough in millimeter-submillimeter astronomy is 3D imaging spectrometry with wide instantaneous spectral bandwidths and wide fields of view. The total optimization of the focal-plane instrument, the telescope, the observing strategy, and the signal-processing software must enable efficient removal of foreground emission from the Earth's atmosphere, which is time-dependent and highly nonlinear in frequency. Here we present \textsf{TiEMPO} : Time-dependent End-to-end Model for Post-process Optimization of the DESHIMA spectrometer. \textsf{TiEMPO}  utilizes a dynamical model of the atmosphere and parametrized models of the astronomical source, the telescope, the instrument, and the detector. The output of \textsf{TiEMPO}  is a timestream of sky brightness temperature and detected power, which can be analyzed by standard signal-processing software. We first compare \textsf{TiEMPO}  simulations with an on-sky measurement by the wideband DESHIMA spectrometer, and find good agreement in the noise power spectral density and sensitivity. We then use \textsf{TiEMPO}  to simulate the detection of the line emission spectrum of a high-redshift galaxy using the DESHIMA 2.0 spectrometer in development. The \textsf{TiEMPO}  model is open source. Its modular and parametrized design enables users to adapt it to design and optimize the end-to-end performance of spectroscopic and photometric instruments on existing and future telescopes.
\end{abstract}


\section{INTRODUCTION}

\begin{figure}
    \centering
    \includegraphics[width=\textwidth]{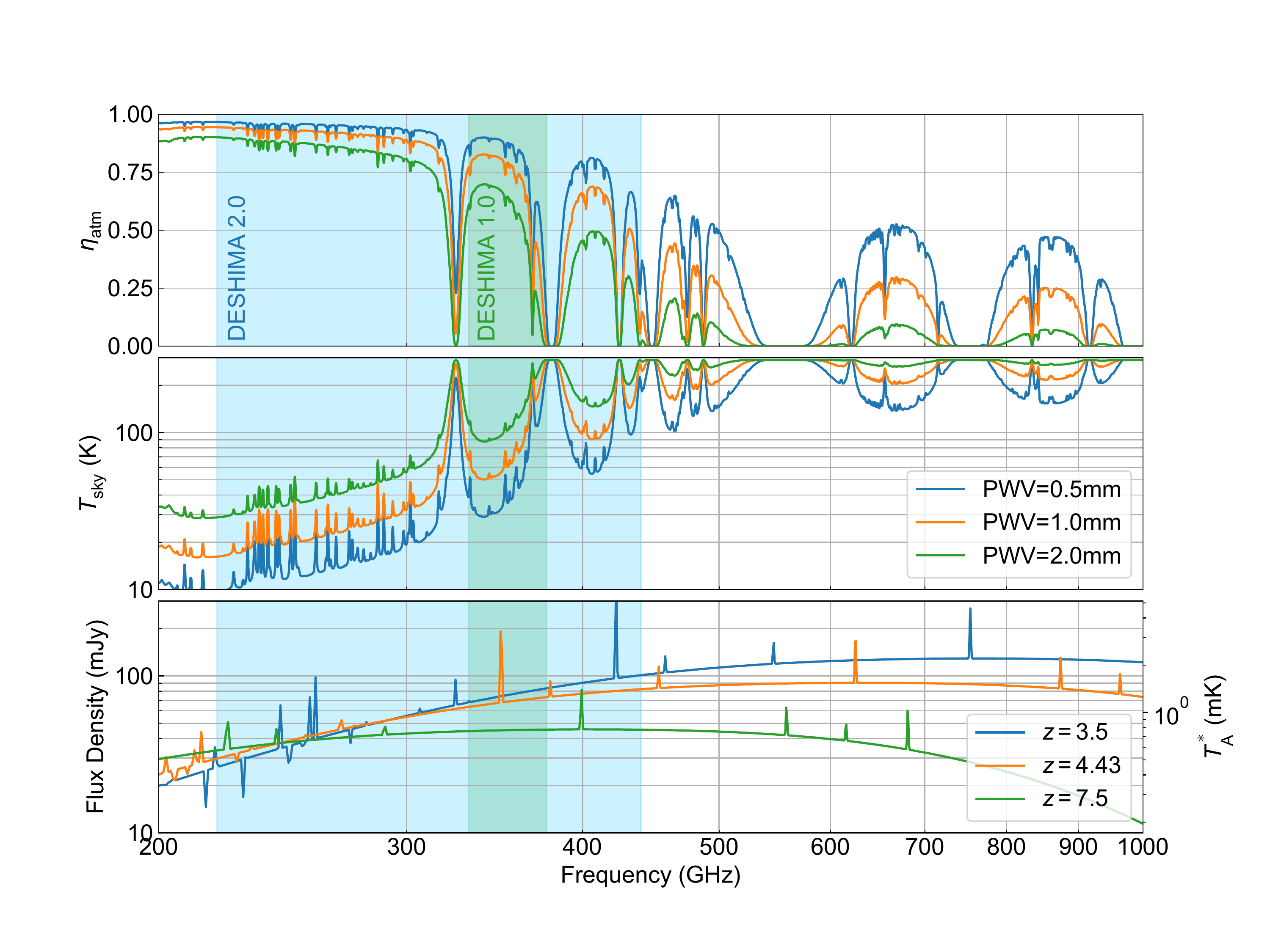}
    \caption{Atmospheric transmittance $\eta_\mathrm{atm}$ (top) and sky brightness temperature $T_\mathrm{sky}$ (middle) at zenith ($\theta=90^\circ$) as functions of frequency, for three values of precipitable water vapor (PWV). The instantaneous frequency coverage of DESHIMA 1.0\cite{Endo2019NatAstron} and the future DESHIMA 2.0 (see Section \ref{section:D2sim}) are indicated by the green and blue shades, respectively. The range of DESHIMA 1.0 is an example of one atmospheric ``window''. DESHIMA~2.0 spans multiple atmospheric windows with absorption-bands in between. (Bottom) \textsf{GalSpec}-simulated spectrum of a galaxy with a far-infrared luminosity of $L_\mathrm{FIR}=10^{13.7}\ L_\odot$, placed at three different redshifts. The spectrum for $z=4.43$ is given as input to \textsf{TiEMPO} in Section~\ref{section:D2sim}. The right vertical axis is a rough indication of the corresponding atmosphere-corrected antenna temperature $T_\mathrm{A}^\ast$, assuming a $\varnothing 10$~m telescope with an aperture efficiency of 0.6.}
    \label{fig:atm}
\end{figure}

The rapidly growing instantaneous bandwidth\cite{Endo2019NatAstron,Endo2019JATIS,Karkare2020,Ade2020,Barrentine2016mu-spec} and field-of-view\cite{Klaassen2020AtLAST,Kawabe2020LST,aravena2019ccatprime} of millimeter-submillimeter (mm-submm) astronomical instruments and telescopes are advantageous not only for collecting more astronomical signal, but also for characterizing and removing the foreground emission of the Earth's atmosphere\cite{Taniguchi2019FMLO}. Even at the best sites for submm astronomy on ground, the brightness temperature of the Earth's atmosphere in the submm range is $\geq$20~K, which can be ${\sim}10^3$--$10^5$ times stronger than the astronomical signal (see Fig.~\ref{fig:atm}). Conventional heterodyne instruments on single-dish telescopes have a typical instantaneous bandwidth of several GHz, which is sufficiently small compared to the atmospheric ``windows'' (the frequency bands over which the atmosphere is relatively transparent). In this narrow-band case, the effect of the atmosphere can often be approximated with a baseline that is linear in frequency. However, the (ultra-)wideband spectrometers in development, such as the Deep Spectroscopic High-redshift Mapper (DESHIMA)\cite{Endo2019NatAstron,Endo2019JATIS,takekoshi2020deshima}, are strongly influenced by the nonlinear frequency dependence of the atmosphere, because they measure across one or even multiple atmospheric windows with strong absorption bands in between. On the one hand, this poses new challenges on the observation and signal-processing techniques to remove the nonlinear atmospheric emission\cite{Taniguchi2019FMLO}. On the other hand, the wideband spectral information of the atmosphere could enable the development and use of advanced signal processing methods for characterizing and ultimately removing the atmospheric component to extract the astronomical signal in a better way. The requirements for applying such techniques are expected to drive the design of future telescopes and focal-plane instrument systems\cite{Klaassen2020AtLAST,Kawabe2020LST,aravena2019ccatprime}.

Here we present \textsf{TiEMPO}, the Time-dependent End-to-end Model for Post-process Optimization of the DESHIMA spectrometer. \textsf{TiEMPO} is a numerical model for simulating wideband submm astronomical observations through the Earth's atmosphere, and produces timestream data that can be fed to data-analysis software\cite{Taniguchi2020decode} as if they were taken with a real instrument operated on a telescope. To account for the nonlinear, dynamic, and inhomogeneous transmittance of the atmosphere, \textsf{TiEMPO}  utilizes the Atmospheric Transmission at Microwave (\textsf{ATM}) model\cite{Pardo2001ATM} to simulate the spectral dependence, and the Astronomical Radio Interferometer Simulator (\textsf{ARIS}) model\cite{Asaki2007VSOP,Asaki2005ARISALMAMemo,Matsushita2017} to simulate the spatial/temporal variations. \textsf{TiEMPO}  is distributed as an open-source \textsf{Python} package and the scripts are available on a public repository\cite{TiEMPOzenodo}, to encourage further use and development by the astronomical community to study cases for different telescope/instrument systems.

\section{The \textsf{TiEMPO} Model}

\subsection{Overview}

\begin{figure}
    \centering
    \includegraphics[width=0.5\textwidth]{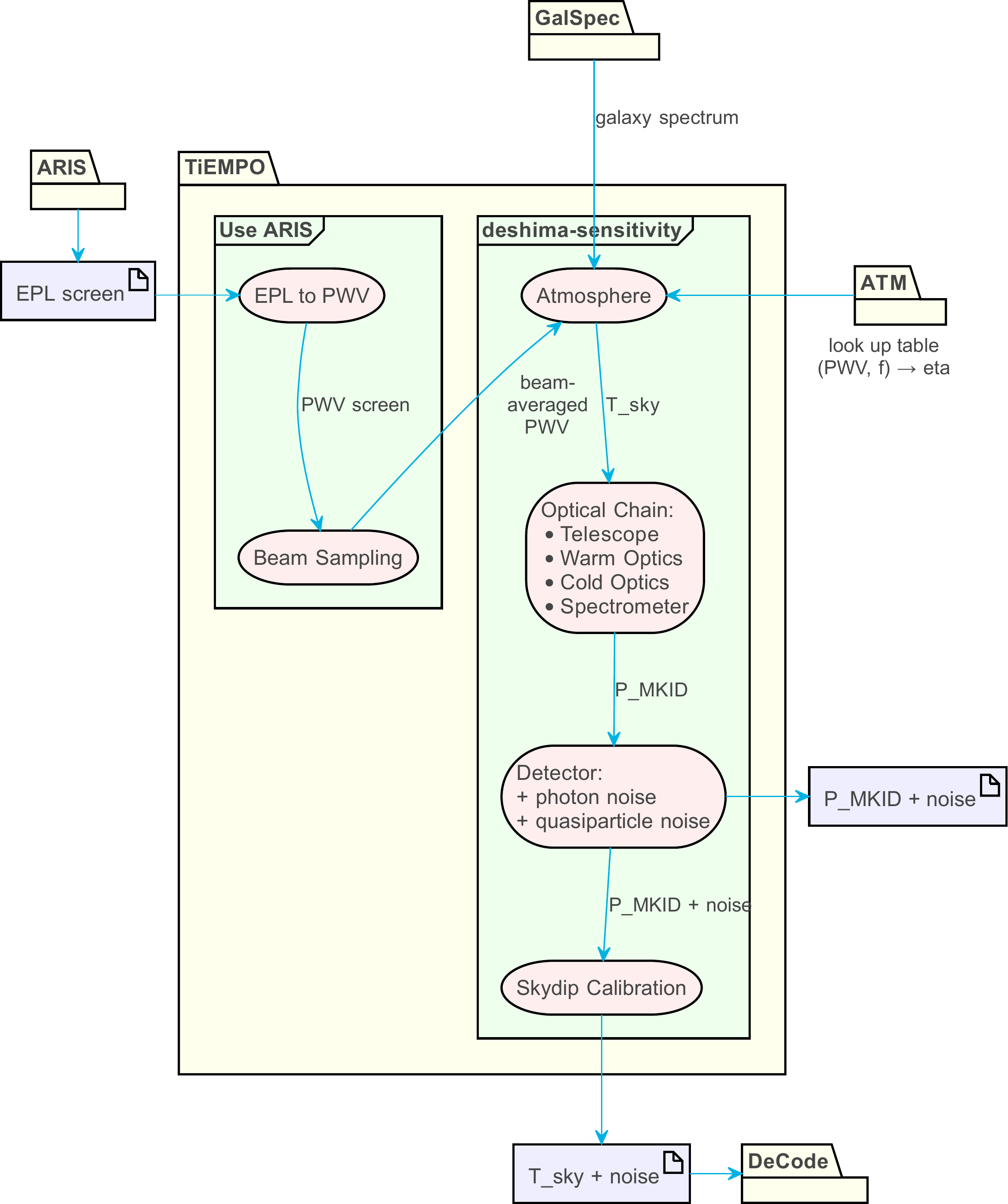}
    \caption{System diagram of \textsf{TiEMPO}, showing each component with its input and output. \textsf{TiEMPO} depends on external packages \textsf{ARIS}\cite{Asaki2007VSOP,Asaki2005ARISALMAMemo}, \textsf{ATM}\cite{Pardo2001ATM}, and \textsf{GalSpec}\cite{galspeczenodo}. \textsf{TiEMPO} outputs calibrated sky brightness temperature $T_\mathrm{sky}$ and detector output $P_\mathrm{MKID}$, which contains photon noise, quasiparticle recombination noise, and atmospheric noise. The output timestream data can be analyzed by post-processing software such as \textsf{De:code}\cite{Taniguchi2020decode}.
    }
    \label{fig:TiEMPO_overview}
\end{figure}

\textsf{TiEMPO} is an end-to-end model, containing models of the astronomical source, the atmosphere, the telescope, the cryogenic instrument optics, and an integrated superconducting spectrometer with microwave kinetic inductance detectors (MKIDs) (see Fig. \ref{fig:TiEMPO_overview}). 
% It can be adapted to model an optical grating spectrometer, or to use another detector type. 
The details of the \textsf{TiEMPO}  model can be found in Refs.\cite{HuijtenBEP,RoelvinkBEP} and the source code is publicly available\cite{TiEMPOzenodo}. In the following we provide an overview of the modules, in the order of signal propagation from the astronomical source to the detector.

\subsection{Dusty High-redshift galaxy}

The galaxy spectrum was created using the \textsf{GalSpec} package\cite{galspeczenodo}, which we distribute as an open-source \textsf{Python} package that can be used outside of \textsf{TiEMPO}.
Our goal is to create a galaxy template that is similar to the types of galaxies we will be trying to detect with DESHIMA, and with which we are able to model the potential future science cases. As such, we have taken an empirical approach to the creation of a galaxy spectrum, combining the continuum shape and spectral line luminosities from recent studies of observed local and high-redshift galaxies. The continuum is based on the two-component modified-black body fit to 24 galaxies at $z > 2$ with Herschel (250, 350 and 500~\textmu m ) and SCUBA-2 (850~\textmu m) fluxes\cite{2018MNRAS.473.1751B,2020MNRAS.494...10B}. Here, we normalize the spectrum to the total far-infrared luminosity by integrating the spectrum from 8 to 1000~\textmu m. The spectral lines are simulated with a more creative approach that can be tailored to specific science goals. Relatively shallow observations, aimed at detecting atomic lines and CO, are simulated using spectral line luminosity scaling relations. Here we use the scaling relations from Ref.\cite{2014MNRAS.438.2547B} for atomic lines and Ref.\cite{2016ApJ...829...93K} for both CO and [CI] lines. The scaling relations of Ref.\cite{2014MNRAS.438.2547B} are based on local star-forming and ultra-luminous infrared galaxies and high-redshift submillimeter galaxies, whereas the scaling relations of Ref.\cite{2016ApJ...829...93K} are based mostly on local galaxies. Deeper observations might resolve more complex molecular lines in both emission and absorption, such as H$_2$O, HCN, HCO$^+$, CH$^+$, NH, NH$_2$, OH$^+$, and HF. These species are only incidentally seen at high redshift (e.g., Ref.\cite{2018Sci...361.1016S}, Berta et al. in prep), and thus we rely on the line detections in the nearby ULIRG Arp~220 to supplement our spectrum for these molecular species\cite{2011ApJ...743...94R}. Here, instead of scaling to the far-infrared luminosity, we scale the line brightness to the observed continuum at the line’s frequency. For this complete galaxy spectrum, we note that the brightness of these spectral lines is a probe of the conditions of the interstellar medium. As such, the line brightnesses (and even the continuum) are known to vary by up to 1~dex from source to source, which must be taken into account when applying for the necessary observation time or detection limits. Throughout this work, we assume a uniform line velocity width of 600~$\mathrm{km\:s^{-1}}$ (full width half maximum), which is found to be the mean for typical SMGs\cite{2013MNRAS.429.3047B} and in line with recent observations of South Pole Telescope and bright Herschel sources\cite{2020ApJ...902...78R,2020MNRAS.496.2372B,2020A&A...635A...7N}.

\subsection{Creation of the Atmosphere Screen}
\label{subsection:atmosphere_screen}
\begin{figure}
    \centering
    \subfloat[]{{\includegraphics[width=8cm]{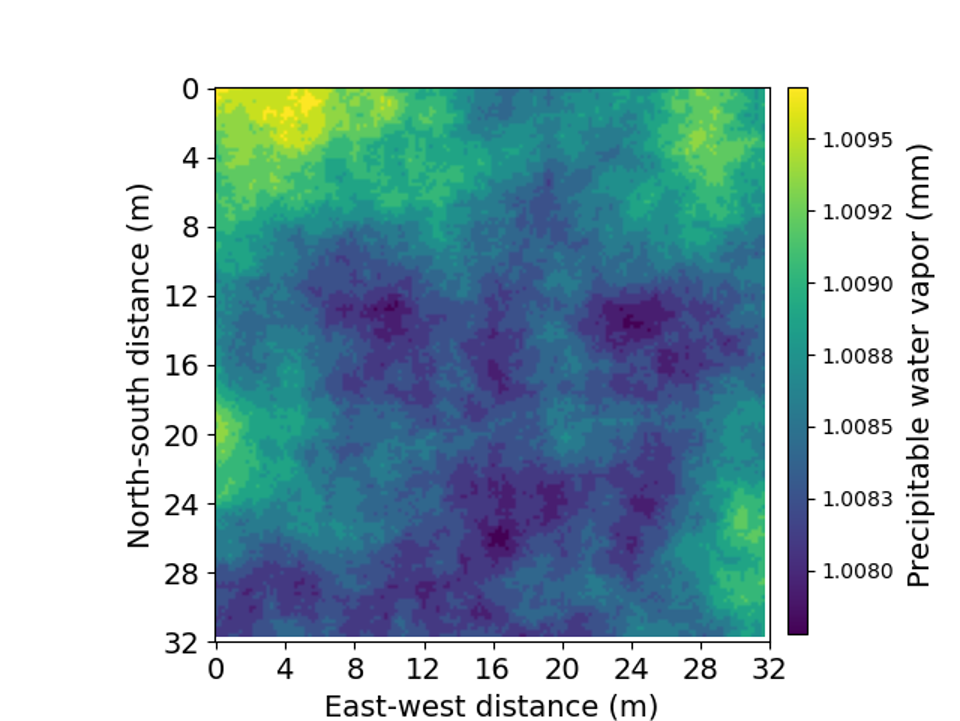}}}
    \qquad
    \subfloat[]{{\includegraphics[width=8cm]{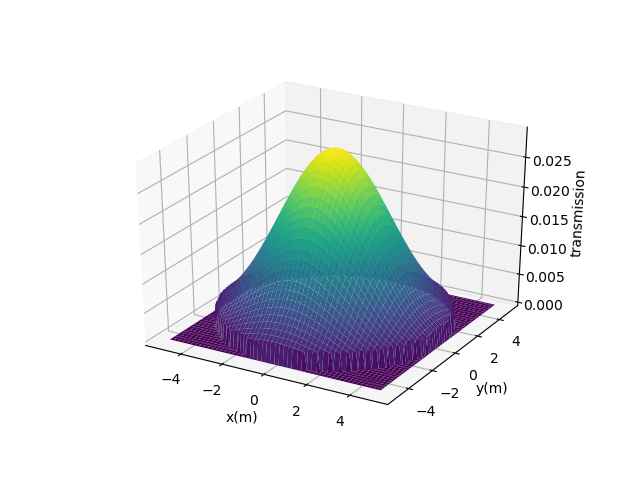}}}
    \caption{ (Left) Colormap of the output of ARIS for a 32~m~$\times$~32~m sky window converted to PWV with Eq.~\ref{Eq:EPL_PWV}. (\href{https://youtu.be/YoI3vVl-ZiU}{\faYoutubePlay \ \textsf{online video}}) (Right) The truncated Gaussian that is used as the telescope beam shape in the model. The volume of the Gaussian is normalized to unity, and it is truncated at a radius of 5~m, where its height is 10\% of its peak height.}
    \label{fig:arismap_and_beam}
\end{figure}

Our goal here is to obtain the line-of-sight transmittance of the atmosphere $\eta_\mathrm{atm}$, which depends on the time $t$ and telescope pointing angle ($\theta$, $\phi$). Here, $\theta$ and $\phi$ are the elevation and azimuth angles of the telescope pointing, respectively. We start from the observation that the mm-submm $\eta_\mathrm{atm}$ at zenith ($\theta = 90^\circ$) is correlated with chiefly one variable, the precipitable water vapor (PWV)\cite{SukulBEP}. Water vapor not only absorbs the submm waves, but also introduces an extra path length (EPL) that is dependent on the line-of-sight PWV\cite{thompson2001interferometry}. \textsf{ARIS}\cite{Asaki2007VSOP,Asaki2005ARISALMAMemo} uses a set of spatial structure functions\cite{dravskikh1979tropospheric} to produce a phase screen, i.e., a two-dimensional map of EPL as shown in the left panel of Fig.~\ref{fig:arismap_and_beam}.

\textsf{TiEMPO} converts an EPL screen to a PWV screen, using the following relation derived from the Smith-Weintraub constants\cite{smith_constants} of EPL and the ideal gas law (see Ref. \cite{HuijtenBEP} for details):
\begin{equation}
\label{Eq:EPL_PWV}
dEPL = 10^{-6} \rho R (k_2 + \frac{k_3}{T})\:dPWV\sim 6.587 \cdot\:dPWV.
\end{equation}
Here,  $k_2 = 70.4 \pm 2.2\ \mathrm{K\ mbar^{-1}}$ and $k_3 = (3.739 \pm 0.012)\cdot 10^5\ \mathrm{K^2\ mbar^{-1}}$ are the Smith-Weintraub constants\cite{smith_constants}, $\rho = 55.4\cdot10^{3}\ \mathrm{mol\ m^{-3}}$ is the number density of molecules in liquid water, $R=8.314\cdot 10^{-2}\ \mathrm{mbar\ m^{3}\ K^{-1}\ mol^{-1}}$ is the gas constant, and $T\sim275$ K is the physical temperature of the atmosphere. Note that $dEPL$ and $dPWV$ are differences from arbitrary mean values of the optical path length and PWV, respectively. In \textsf{TiEMPO}  the user can specify a mean PWV, around which the PWV fluctuates according to the ARIS-modeled EPL and Eq.~\ref{Eq:EPL_PWV}. The mean PWV can be set to a constant, or a vector that represents a slowly changing weather condition. The created PWV screen moves spatially in one direction, at the user-specified wind velocity, assuming that the structure of phase fluctuations is invariant when the atmosphere moves with the wind\cite{gurvich1967atmospheric}. See the left panel of Fig.~\ref{fig:arismap_and_beam} and the online animation for an example of the moving atmosphere created in \textsf{TiEMPO}.

\subsection{Sampling of the Atmosphere by the Telescope Near-Field Beam}

The water vapor in the atmosphere above the Atacama Desert is contained mostly in the layer up to $\sim$1 km from ground\cite{giovanelli2001optical}, which is well within the near field of the telescope. Therefore, the beam pattern at this height has a similar pattern to the power distribution over the primary mirror of the telescope. For simulating DESHIMA on ASTE using \textsf{TiEMPO}, we have assumed a Gaussian power pattern as shown in the right panel of Fig.~\ref{fig:arismap_and_beam}, which drops to $-10$ dB at the telescope radius of 5 m. The PWV map created in Subsection~\ref{subsection:atmosphere_screen} is filtered with this beam pattern, so that the received power from the atmosphere is a weighted average within the beam. The user may include an arbitrary beam pattern in \textsf{TiEMPO}, when detailed information is available from the design or measurement.

\subsection{The Far-Field Beam of the Telescope}

The far-field telescope beam is modeled by two properties\cite{wilson_tools_2009}: the main beam solid angle $\Omega_\mathrm{MB}$ and the main beam efficiency $\eta_\mathrm{MB}$. $\Omega_\mathrm{MB}$ represents the solid angle of the beam excluding the side lobes. $\eta_\mathrm{MB}$ is the fraction of the beam contained in the main beam, out of the total reception pattern. The (total) beam solid angle, with the side lobes included, is then given by
\begin{equation}
    \Omega_\mathrm{A} = \frac{\Omega_\mathrm{MB}}{\eta_\mathrm{MB}}.
\end{equation}
We use the beam solid angle to define the effective aperture area:
\begin{equation}
    A_{e} = \frac{\lambda^{2}}{\Omega_{A}},
\end{equation}
where $\lambda$ is the wavelength.
Now, we can express the aperture efficiency $\eta_\mathrm{A}$ as
\begin{equation}
    \eta_\mathrm{A} = \frac{A_\mathrm{e}}{A_\mathrm{p}} = \frac{\eta_\mathrm{MB}}{\Omega_\mathrm{MB}} \frac{\lambda^{2}}{A_\mathrm{p}},
    \label{eq:eta_A}
\end{equation}
where $A_\mathrm{p}$ is the physical area of the telescope primary mirror. 
From these quantities, the single-mode power density (in $\mathrm{W\:Hz^{-1}}$) of the astronomical source is calculated as
\begin{equation}
    P_f = \frac{1}{2} F_f A_\mathrm{e}, 
\end{equation}
where $F_{f}$ denotes the flux density in $\mathrm{W\ m^{-2}\ Hz^{-1}}$ ($=10^{26}\ \mathrm{Jy}$) and the factor $1/2$ compensates for the fact that the flux density is calculated using two polarizations, but the power density that \textsf{TiEMPO} calculates is for single-polarization assuming the coupling of the signal to a single-mode (on-chip) antenna and transmission line\cite{Endo2019JATIS}. 

\subsection{Radiative Transfer}

\label{sec:desim}

For calculating the single-mode radiation transfer from the astronomical source to the detector, a subset of the \textsf{deshima-sensitivity} software\cite{deshima-sensitivity} was used. Each component in the optical chain is modeled with a black body power density and a transmission factor $\eta_i$. The single-mode power density of a black body is equivalent to the Johnson-Nyquist noise, and is given by
\begin{equation}
    P_f = \frac{hf}{e^{\frac{hf}{k_\mathrm{B}T}} - 1}.
    \label{eq:Johnson-Nyquist}
\end{equation}
Here, $h$ is the Planck constant, $f$ is the frequency, $k_\mathrm{B}$ is the Boltzmann constant and $T$ is the physical temperature of the emitter\cite{nyquist1928thermal}. The spectral power before the detector is computed by cascading the radiation transfer of each component as:
\begin{equation}
    P_{f,\mathrm{out}} = \eta_i \: P_{f,\mathrm{in}} + (1-\eta_i) P_{f,\mathrm{medium}}, 
    \label{eq:rad_trans}
\end{equation}
where $P_{f,\mathrm{out}}$ is the power density of the radiation that comes out of the component, $P_{f,\mathrm{in}}$ is the power density of the radiation going in, $\eta_i$ is the transmittance of the medium, and $P_{f,\mathrm{medium}}$ is the power density of the medium. 

\subsection{Spectrometer}

\textsf{TiEMPO} is able to model any direct-detection (imaging-)spectrometer that couples the wideband input power into one or more spectral channels. Examples include integrated filterbank spectrometers that use a filterbank\cite{Endo2019JATIS,Endo2019NatAstron,Karkare2020} or an integrated grating\cite{Ade2020,Barrentine2016mu-spec}, and optical grating spectrometers. Currently, \textsf{TiEMPO}  takes three spatial pixels to simulate position-switching observations, but the pixel count can be increased to model multi-pixel imaging arrays. If the number of spectral channels per pixel is set to unity, then the model can represent a monochromatic imaging camera. 

\textsf{TiEMPO} can import the spectral response of each detector, obtained from the design or a measurement. Here we have assumed a simple Lorentzian-shaped spectral transmission, which is a good approximation for a filterbank channel\cite{Endo2019JATIS}, or a detector behind an optical (or substrate-integrated) grating\cite{Barrentine2016mu-spec}. The frequency dependence was implemented by dividing the frequency range of 210--450~GHz (10~GHz wider on each side than the nominal DESHIMA~2.0 band, to take into account power coupling from outside of the band) into 1500 bins. The resulting efficiency is used to compute the power density with the radiation transfer equation, Eq.~\ref{eq:rad_trans}. Finally, the power in each bin is calculated as
\begin{equation}
    P_{\mathrm{bin}\:i} = \Delta f \: P_{f,\mathrm{bin}\:i}.
    \label{eq:binning_power}
\end{equation}
Note that the $P_{\mathrm{bin}\:i}$ in Eq.~\ref{eq:binning_power} is the expected value of the power. To calculate the frequency-integrated power detected by each detector at each moment $t$, $P_\mathrm{MKID}(t)$, we must consider noise (see Subsection \ref{sec:NEP}).

\subsection{Detected power and noise}
\label{sec:NEP}
The best possible sensitivity of a pair-breaking detector like an MKID is set by the photon noise and quasiparticle recombination noise. The commonly-used narrow-band approximation for the noise equivalent power (NEP) limited by photon- and recombination-noise is given by\cite{Endo2019NatAstron} 
\begin{equation}
\label{eq:NEPph}
\begin{split}
NEP_\mathrm{{MKID}} &= \sqrt{2\overline{P_\mathrm{MKID}}(hf+\overline{P_\mathrm{MKID}}/\Delta f)+4\Delta_\mathrm{Al} \overline{P_\mathrm{MKID}}/\eta_\mathrm{pb}}.
\end{split}
\end{equation}
Here, $\overline{P_\mathrm{MKID}}$ is the expected value of the power \textit{absorbed} by the MKID, $\Delta_\mathrm{Al}=188\ \mathrm{\mu eV}$ is the superconducting gap energy of aluminium, and $\eta_\mathrm{pb}\sim 0.4$ is the pair-breaking efficiency\cite{2014SuScT..27e5012G}. In \textsf{TiEMPO} we use the more general, integral form of the NEP\cite{Zmuidzinas2003Photon} to take into account a frequency-dependent optical efficiency over a wide bandwidth, for each detector of the spectrometer. Because the fluctuation in energy in different spectral bins are uncorrelated\cite{Richards1994}, we can calculate the standard deviation in power for each spectral bin per sampling rate ($1/f_\mathrm{sampling}$) from
\begin{equation}
    \sigma = NEP_\mathrm{MKID}\sqrt{\frac{1}{2} f_\mathrm{sampling}},
    \label{eq:st_dev}
\end{equation}
and add those together to obtain the combined detector output:
\begin{equation}
    P_\mathrm{MKID} = \sum^\mathrm{\#bins}_{i=1} P_{\mathrm{bin\:}i,\mathrm{\:with\:noise}}
    \label{eq:binning_sum}
\end{equation}
Note that this is equivalent to calculating the detector NEP directly from the integral:
\begin{equation}
\label{eq:NEPinteg}
\begin{split}
NEP_\mathrm{MKID}^2 = \int 2P_f(hf +  P_f) + 4\Delta_\mathrm{Al} P_f/\eta_\mathrm{pb}\:df.
\end{split}
\end{equation}
The integration over a wide bandwidth taking the filter spectral transmission into account is especially relevant for spectral channels that are near strong emission lines and absorption bands of the atmosphere\cite{takekoshi2020deshima}.

\subsection{Sky temperature Calibration}
\label{sec:conv_PtoT}
After we have computed the noise-added power that is measured by the MKIDs, we want to relate this back to the original signal from the sky. We do this by expressing the received power in sky brightness temperature $T_\mathrm{sky}$: the physical temperature of a black body that would have the same intensity as the semitransparent sky\cite{Endo2019NatAstron}. To this end we take the Johnson-Nyquist formula given by
\begin{equation}
    T_\mathrm{sky} = \frac{h f}{k_\mathrm{B}\:\ln(\frac{hf}{P_f}+1)}.
    \label{eq:Planck}
\end{equation}

In order to relate the MKID power $P_\mathrm{MKID}$ to $T_\mathrm{sky}$, \textsf{TiEMPO}  internally performs a skydip simulation \cite{takekoshi2020deshima} using the \textsf{deshima-sensitivity}\cite{deshima-sensitivity} script. A skydip is a series of measurements in which the telescope `dips' from a high elevation (pointing at zenith) to a low elevation (pointing almost horizontally). When the elevation is lower, the telescope looks through a thicker layer of atmosphere, increasing the opacity, and hence the power and the sky temperature, allowing us to construct a relationship between the two. In our simulations, we use elevation values in the range of $\theta=20^\circ$--90$^\circ$. The power and sky temperature data are interpolated for each channel and saved in the model. \textsf{TiEMPO}  reuses these interpolation curves, so that they only need to be created once. For further details on the numerical skydip calibration, see Ref.\cite{HuijtenBEP}.

\section{Comparing a \textsf{TiEMPO} simulation with on-sky DESHIMA 1.0 measurements}

In order to verify the \textsf{TiEMPO}  model, we made a simulation of an atmosphere observation using input parameters that resemble a real measurement done with the DESHIMA spectrometer on the ASTE telescope\cite{ASTE_Ezawa2004}. DESHIMA is an integrated superconducting spectrometer with MKID detectors. The first generation of DESHIMA, which we will hereafter call DESHIMA~1.0\cite{Endo2019NatAstron}, has an instantaneous band of 332--377~GHz, with a frequency spacing of $f/\Delta f\sim 380$. (The half-power bandwidth of each filter is $f/\Delta f\sim 300$ on average\cite{Endo2019JATIS}.) ASTE is a 10-m Cassegrain reflector located on the Pampa la Bola plateau of the Atacama Desert in northern Chile, at an altitude of 4860~m. DESHIMA~1.0 was operated on ASTE during October--November 2017\cite{Endo2019NatAstron}. The measured response of the MKIDs was converted to sky brightness temperature $T_\mathrm{sky}$ using a skydip calibration method explained in detail in Ref. \cite{takekoshi2020deshima}.

\begin{figure}
    \centering
    \includegraphics[width=0.6\textwidth]{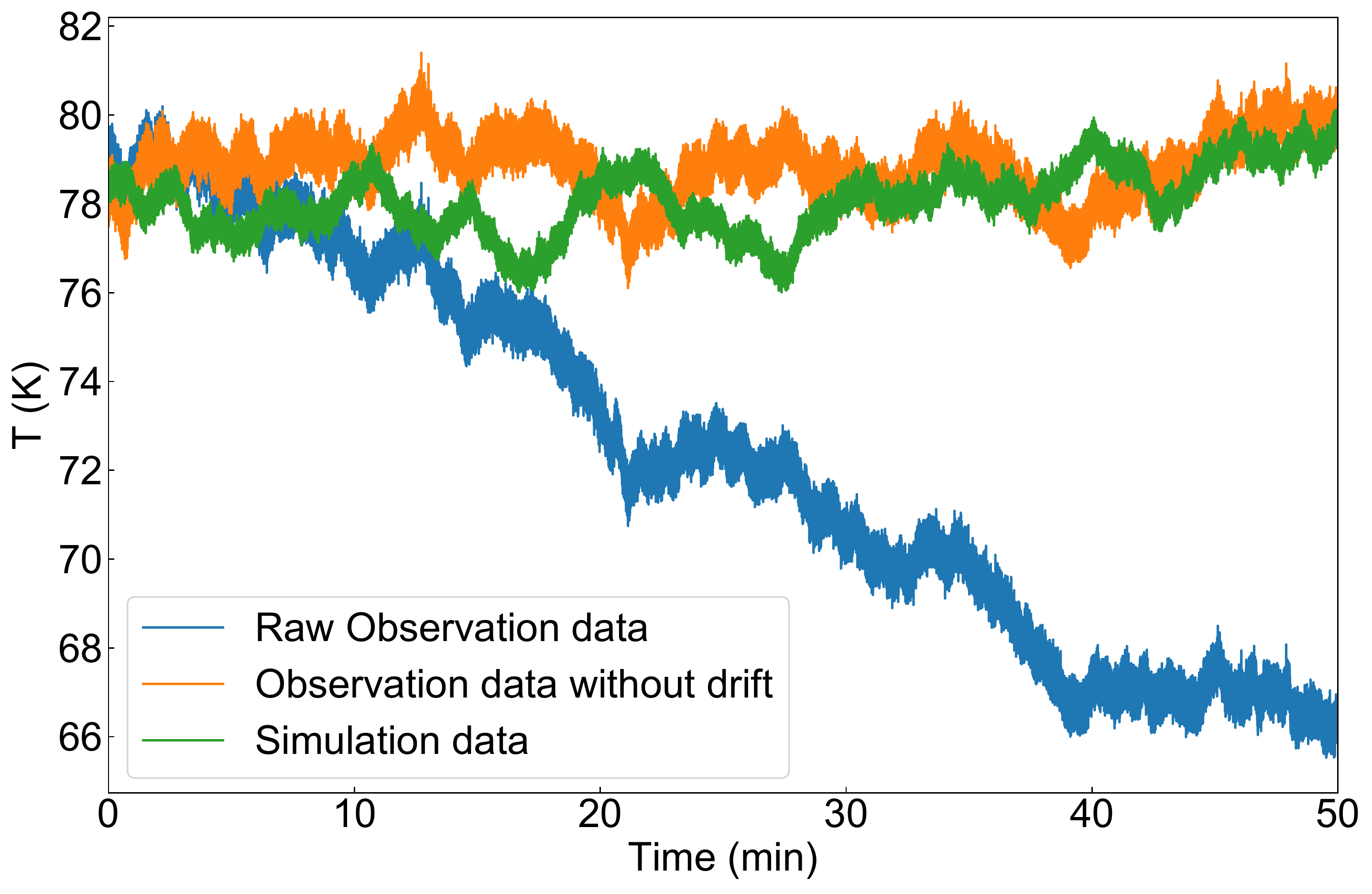}
    \caption{Time stream of $T_\mathrm{sky}$, for DESHIMA 1.0 observation (blue) and TiEMPO simulation (green). For better comparison, the orange trace shows the observation data but with its linear drift subtracted.}
    \label{fig:timestream}
\end{figure}

We use the data taken from a measurement on November 17th 2017, in which the telescope was pointed close to zenith ($\theta$ = 88$^\circ$) for 3000~s. The PWV measured with the radiometer of the Atacama Large Millimeter-submillimeter Array (ALMA) was 1.7~mm at the beginning of the measurement, corresponding to a 350~GHz sky brightness temperature of $\sim$78~K. In Fig.~\ref{fig:timestream} we show the time-evolution of $T_\mathrm{sky}$ taken with the 350~GHz channel of DESHIMA~1.0 (blue curve). The DESHIMA measurement indicates that the PWV dropped continuously over the course of the measurement, from $\sim$1.7~mm to $\sim$1.3~mm. 

The \textsf{TiEMPO}-simulated time trace of $T_\mathrm{sky}$ for the 351~GHz channel is compared to that of the DESHIMA~1.0 measurement in Fig.~\ref{fig:timestream}. An obvious difference is that the measured curve steadily decreases with time, but the simulated curve stays at $T_\mathrm{sky}\sim78$~K. This is because the \textsf{TiEMPO}-simulation was set to a constant time-averaged PWV of 1.72~mm. To aid visual comparison of the fluctuations we have included the orange curve in Fig.~\ref{fig:timestream}, which is the observed data but with its linear component subtracted. If we now compare the green (\textsf{TiEMPO}) and orange (flattened observation) traces, there appears to be a fairly good agreement. Both curves show two types of fluctuations that are behaving similarly: the low-frequency and large amplitude fluctuations are due to atmospheric noise, whereas the high-frequency and small amplitude fluctuations are due to photon noise. 

\begin{figure}
    \centering
    \includegraphics[width=0.6\textwidth]{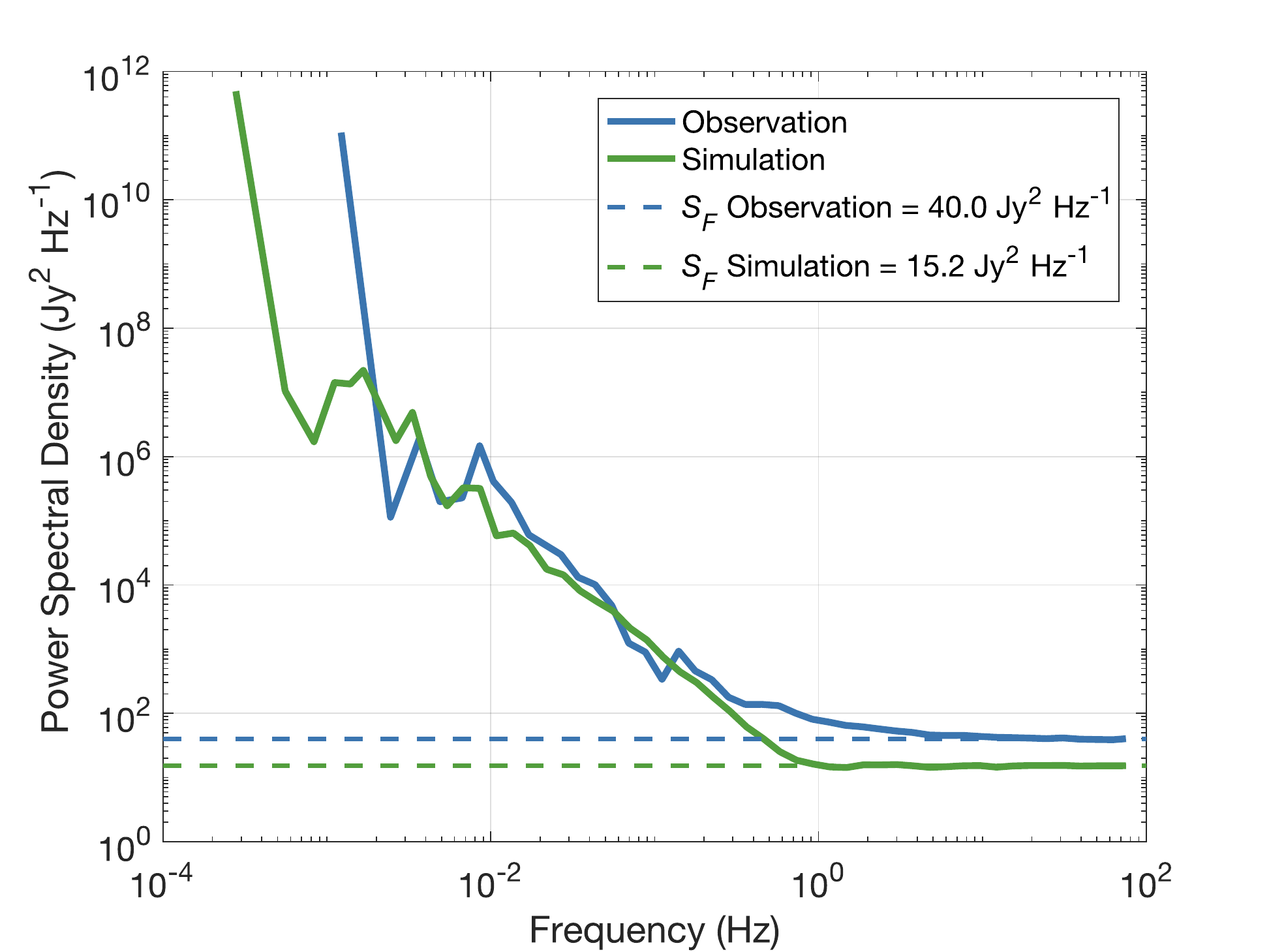}
    \caption{Noise power spectral density (PSD) of the simulated and measured sky flux density. The dashed lines indicate the average level at above 10~Hz where the white photon noise is dominating.}
    \label{fig:skyPSD}
\end{figure}

To compare the noise in the simulation and measurement more quantitatively, we have taken the power spectral density (PSD) of the time traces. Before taking the PSD, we have converted $T_\mathrm{sky}$ to flux density, by
\begin{equation}\label{eq:Fv_Johnson}
    F_f(T_{\mathrm{sky}}) = \frac{2 \Omega_{\mathrm{MB}}}{\eta_{\mathrm{MB}} \lambda^2 \eta_{\mathrm{atm}}} \frac{hf}{e^{\frac{hf}{kT_{\mathrm{sky}}}}-1}.
\end{equation}
In Fig.~\ref{fig:skyPSD} we show the resulting PSDs for the simulation and measurement. At $<$1~Hz, the PSD curves exhibit the $1/f$-noise generated by the atmospheric fluctuations. At $> 1$ Hz, the PSDs flatten out to the white photon noise level. This form of PSD is often seen in mm-submm observations\cite{Monfardini:2011ii}. The similarity between the two PSDs indicates that TiEMPO is able to simulate both the atmospheric and photon noise. The photon noise level $S_F$ of the observation is 2.6 times higher than that of the simulation, which is consistent with the $\times$1.6 (${\sim}\sqrt{2.6}$) difference in the photon noise amplitude in the time stream shown in Fig.~\ref{fig:timestream}. This difference suggests a difference in optical coupling to the sky, which is not surprising because the simulation assumed a constant optical efficiency and filter half-power bandwidth ($f/\Delta f=300$) for all spectral channels, but the real DESHIMA~1.0 instrument has channel-to-channel variations in these quantities\cite{Endo2019JATIS}. A more detailed analysis taking into account the measured characteristics of each channel finds a good agreement in the photon-noise level\cite{MarthiInternship}.

\begin{figure}
    \centering
    \includegraphics[width=0.6\textwidth]{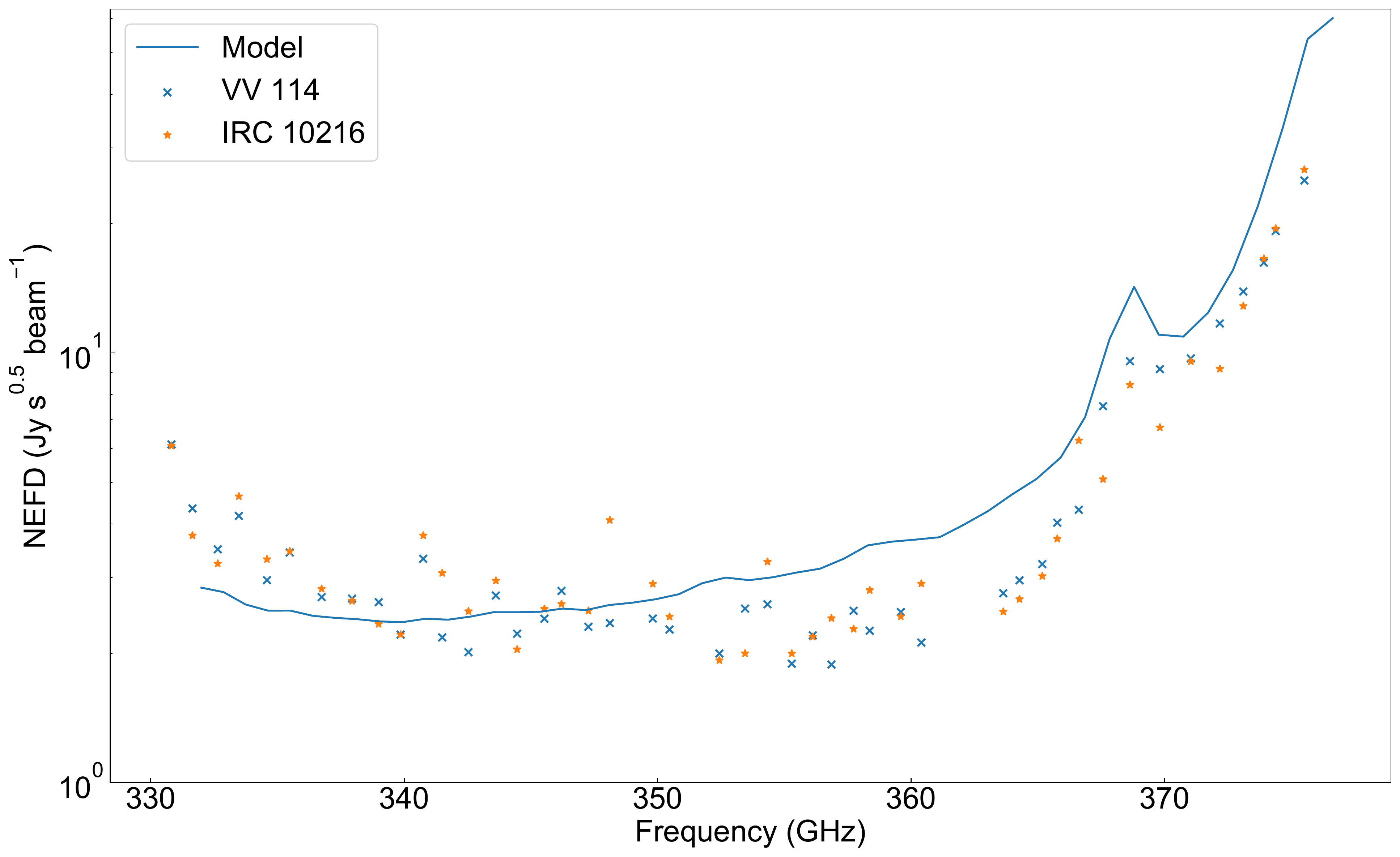}
    \caption{Noise equivalent flux density (NEFD) calculated from the flat photon-noise level ($S_F$) of the \textsf{TiEMPO}-simulated data of Fig.~\ref{fig:skyPSD}, compared to the NEFD of DESHIMA~1.0 based on actual detection of astronomical emission lines, from the luminous infrared galaxy VV~114 and the post-asymptotic giant branch star IRC$+$10216\cite{Endo2019NatAstron}.
    }
    \label{fig:NEFD}
\end{figure}

Using the flat photon-noise level $S_F$ calculated from the \textsf{TiEMPO}-simulated PSD, we have calculated the noise equivalent flux density as an indicator of the system sensitivity by $NEFD=\sqrt{S_F/2}$. We plot the NEFD of all 49 simulated channels in Fig.~\ref{fig:NEFD}, together with the measured NEFD of DESHIMA~1.0 on ASTE based on actual measurements of astronomical line detection\cite{Endo2019NatAstron}. Considering the above-mentioned simplified filter model, as well as the measurements taken across nights with different atmospheric conditions\cite{Endo2019NatAstron}, the agreement is  good. 

In summary of this section, the output of \textsf{TiEMPO} resembles the on-sky measurement of DESHIMA~1.0, in both time domain and frequency domain. The end-to-end system sensitivity inferred from the simulation is in good agreement with the actual measurement of astronomical sources performed by DESHIMA~1.0 on ASTE.

\section{TiEMPO Simulation of observing a high-redshift galaxy with DESHIMA~2.0 on ASTE}
\label{section:D2sim}

As an example application of \textsf{TiEMPO}, we simulate the observation of a luminous dusty galaxy ($L_\mathrm{FIR}=10^{13.7}\:L_\odot$) at redshift $z=4.43$ and velocity width 600~$\mathrm{km\:s^{-1}}$ using the DESHIMA~2.0 spectrometer on the ASTE telescope. DESHIMA~2.0 is an upgrade of DESHIMA~1.0, which is currently under development\cite{pascual_laguna_wideband_2019}. The target instantaneous frequency coverage is 220--440~GHz, with a frequency spacing and half-power channel bandwidth of $f/\Delta f = 500$. The system will include a rotating mirror in the cabin optics that enables position switching on the sky at a rate of up to 10~Hz. Assuming the use of this beam chopper, we have simulated a so-called ``ABBA'' chop-nod observing technique\cite{Archibald2002,RoelvinkBEP} with a beam-chopping frequency of 10~Hz between on-source and off-source positions, and a nodding cycle of 60~s to subtract the atmospheric emission from the spectrum. The total simulated observation time was 1~hour. The input spectrum of the galaxy was simulated using \textsf{GalSpec}. The telescope elevation angle was kept constant at 60$^\circ$, and the weather condition was set to: mean PWV = 1.0~mm; root-mean-square fluctuation of the EPL $\sigma_\mathrm{EPL}=50$~\textmu m; wind velocity $= 9.6\:\mathrm{m\:s^{-1}}$. 

\begin{figure}[t]
    \centering
    \includegraphics[width=1\textwidth]{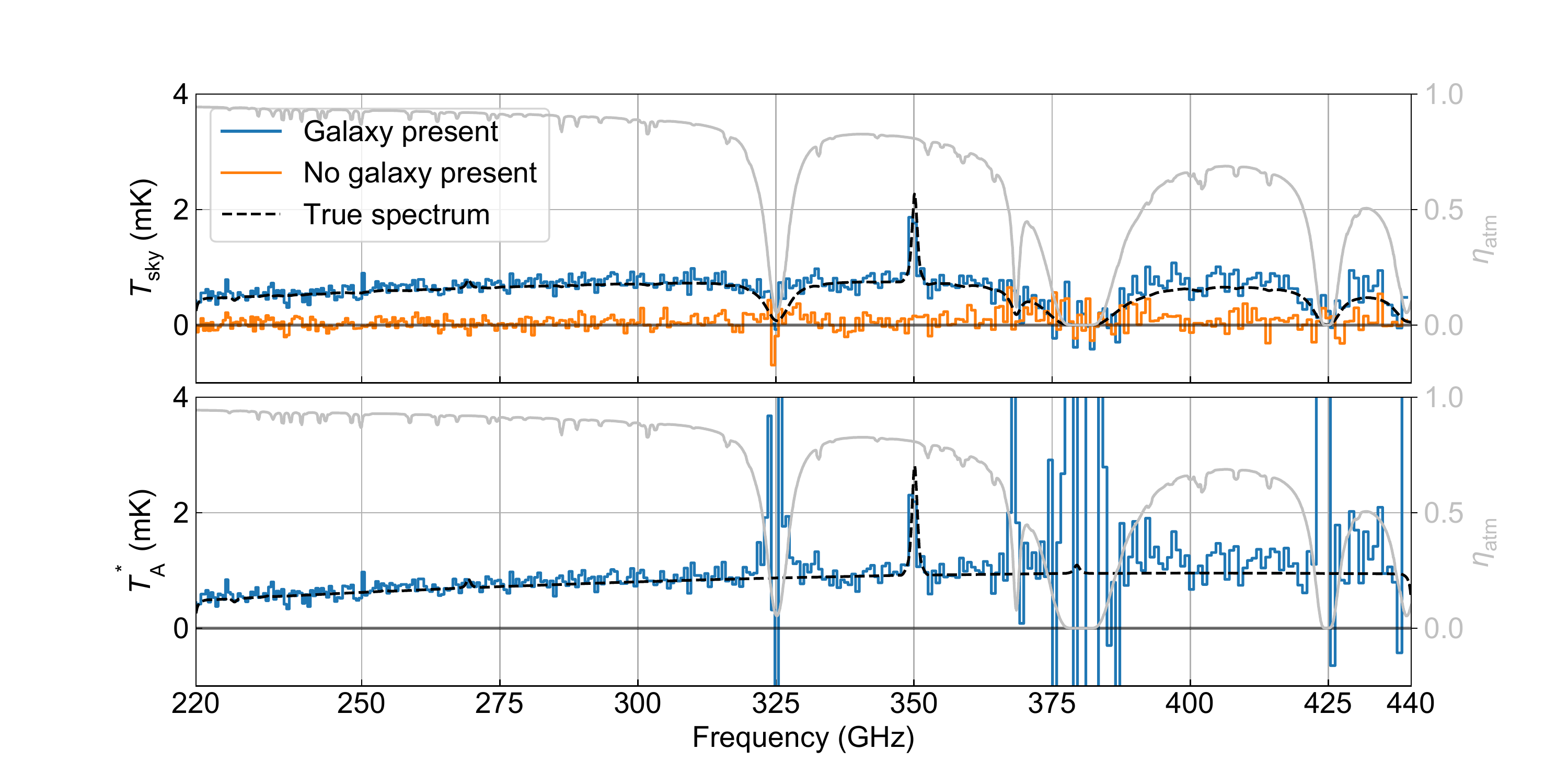}
    \caption{
    \textsf{TiEMPO}-simulated detection of a high-redshift dusty galaxy with the upcoming DESHIMA~2.0 instrument on ASTE. The galaxy spectrum was simulated using \textsf{GalSpec}, with input parameters as follows: far-infrared bolometric luminosity $10^{13.7}\ L_\odot$; redshift $z=4.43$; velocity width 600~$\mathrm{km\ s^{-1}}$. The total observing time was 60~min, out of which half was pointing on-source. \textbf{(Top)} The difference in sky temperature $T_\mathrm{sky}$ between on-source and off-source, obtained with the ABBA chop-nod method. The blue spectrum is the result of placing a galaxy at the on-position, whilst the orange spectrum is the result of no galaxy being present. The dashed curve is the expected spectrum of the galaxy, multiplied by the atmospheric transmittance $\eta_\mathrm{atm}$ (gray) and smoothed with a Lorentzian window of $f/\Delta f=500$ to account for the resolving power of the spectrometer. \textbf{(Bottom)} The atmosphere-corrected antenna temperature $T_\mathrm{A}^* = T_\mathrm{sky}/\eta_\mathrm{atm}$. The blue curve shows the simulated spectrum, whilst the dashed curve is what is expected directly from the input galaxy spectrum. 
    }
    \label{fig:D2spectrum}
\end{figure}

The resulting spectrum after applying the ABBA atmosphere removal scheme is presented in Fig.~\ref{fig:D2spectrum}. The top panel shows the spectrum in $T_\mathrm{sky}$, that is before correcting for atmospheric absorption. The spectrum shows the detection of the redshifted [CII] line at 350~GHz, and the dust continuum emission. In the same figure, we show that a reference simulation without a galaxy yields a zero-centered spectrum as expected. Dividing the $T_\mathrm{sky}$-spectrum by the frequency-dependent atmospheric transmittance $\eta_\mathrm{atm}$ yields the spectrum presented in the bottom panel, where the scale is now atmosphere-corrected antenna temperature $T_\mathrm{A}^*$. In this way, TiEMPO is able to simulate end-to-end observations of future instruments and forecast their scientific products. The \textsf{TiEMPO} data can also be used to optimize observing strategies and signal-processing techniques before the instrument is deployed. 

\section{Conclusion and Future Prospects}

We have presented the \textsf{TiEMPO} model and verified its applicability by comparing its output to on-sky measured data, and simulating the operation of a future instrument. The \textsf{TiEMPO}  model is highly parametrized and modular, so that it can be adjusted to different observation techniques, telescopes, and instruments. This can be done by either simply adjusting the input parameters, or by relatively simple modifications of the \textsf{Python} code. For example, some of the authors have adapted \textsf{TiEMPO}  to simulate scan-mapping observations\cite{ZaalbergBEP}, or included excess detector noise\cite{MarthiInternship}. \textsf{TiEMPO}  can also import arbitrary frequency-dependent transmission curves from models or measurements, to replace the Lorentzian filter transmission used in this article\cite{MarthiInternship}. The time-dependent telescope elevation can be given as a user-specified vector. If the detector is not a pair-breaking type (e.g., superconducting transition-edge sensors), then the recombination noise term can be omitted in Eq.~\ref{eq:NEPph}.

It would seem especially interesting to use \textsf{TiEMPO} for the design and optimization of large mm-submm telescopes, such as the Large Submillimeter Telescope\cite{Kawabe2020LST} and Atacama Large Aperture Submillimeter Telescope (AtLAST)\cite{Klaassen2020AtLAST}, as well as for optimizing instruments and observing techniques on existing large telescopes like the Large Millimeter Telescope (LMT)\cite{Hughes2016LMT}. These telescopes have diameters in the range of 30--50~m, so they sample a larger column of atmosphere that contains a larger number of patches of water vapor that can influence the noise behavior. The combination of \textsf{TiEMPO}  and \textsf{ARIS} can already simulate observations with telescopes of these sizes, in combination with the wideband direct detection imaging spectrometers that are considered as candidates for the future instruments. Note that the current \textsf{TiEMPO} simulates only the transmittance of the atmosphere, and it does not model the wavefront distortion caused by the dynamical and spatially dependent EPL. 
Since \textsf{ARIS} provides an EPL screen, this would be an interesting direction for future development.

\section{Code Availability}

\textsf{TiEMPO}\cite{TiEMPOzenodo}, \textsf{GalSpec}\cite{galspeczenodo}, \textsf{deshima-sensitivity}\cite{deshima-sensitivity}, and \textsf{De:code}\cite{Taniguchi2020decode} are all available as \textsf{Python} packages and distributed on a public repository.

\acknowledgments

We would like to thank Yoshiharu Asaki for providing us with knowledge about the Atacama atmosphere and the \textsf{ARIS} model, including multiple upgrades of \textsf{ARIS} to enable the use of large phase screens for \textsf{TiEMPO}. We would also like to thank Henry Kool for the optimization of the computation server used for this study. Most of the development and validation of \textsf{TiEMPO} was carried out as the TU Delft Bachelor End Projects of EH and YR. SAB completed the development and published the \textsf{TiEMPO} package as part of her TU Delft Master End Project. AE was supported by the Netherlands Organization for Scientific Research NWO (Vidi grant n$^\circ$ 639.042.423). JJAB was the supported by the European Research Counsel ERC (ERC-CoG-2014 - Proposal n$^\circ$ 648135 MOSAIC). YT was supported by the Japan Society for the Promotion of Science JSPS (KAKENHI Grant n$^\circ$ JP17H06130).  The ASTE telescope is operated by National Astronomical Observatory of Japan (NAOJ).

% References
\bibliography{ms.bib} % bibliography data in report.bib
\bibliographystyle{spiebib} % makes bibtex use spiebib.bst

\end{document}